\documentclass[preprint,showpacs,preprintnumbers,amsmath,amssymb]{revtex4}

\usepackage{graphicx}
\usepackage{dcolumn}
\usepackage{bm}

\begin{document}

\title{Dynamical Mean-Field Theory for Molecules and Nanostructures}
\author{V.~Turkowski$^{a,b}$,\altaffiliation{Corresponding author, e-mail address: vturkows@ucf.edu} 
 A.~Kabir$^{a}$, N.~Nayyar$^{a}$,
and Talat~S.~Rahman$^{a,b}$}

\affiliation{$^{a}$ Department of Physics,University of Central Florida, Orlando, FL 32816\\
 $^{b}$ NanoScience and Technology Center,
University of Central Florida, Orlando, FL 32816}

\date{\today}

\begin{abstract}
Dynamical Mean-Field Theory (DMFT) has established itself as a reliable and well-controlled approximation to study correlation effects in bulk solids and also two-dimensional systems. 
In combination with standard density-functional theory (DFT) it has been successfully applied to study materials in which localized electronic states play an important role. There are several evidences that for extended systems this DMFT+DFT approach is more accurate than the traditional DFT+U approximation, particularly because of its ability to take into account dynamical effects, such as the time-resolved double occupancy of the electronic orbitals.   It was recently shown that this approach can also be successfully
applied to study correlation effects in nanostructures.  Here, we present a brief review of the recently proposed generalizations of the DFT+DMFT method. In particular, we discuss in details our recently proposed DFT+DMFT approach to study the  magnetic properties
of nanosystems\cite{Turkowski}  and  present its application to small (up to five atoms) Fe and FePt clusters. We demonstrate that being  a mean-field approach, DMFT produces meaningful results even for such small systems. We compare our results with those obtained using DFT+U and find that, as in the case of bulk systems, the latter approach tends to overestimate correlation effects in nanostructures. Finally, we discuss possible ways to farther improve the nano-DFT+DMFT approximation and to extend its application  to molecules and nanoparticles on substrates 
and to nonequilibrium phenomena.
\end{abstract}

\pacs{75.10.-b, 75.75.-c, 75.40.Mg, 75.50.Bb, 75.50.Cc}

\maketitle

\section{Introduction} 

The theoretical description of solid state systems which include localized electronic states (usually in the case of transition metal (TM) atoms with unfilled shells of d-electrons) remains one of the most difficult problems of condensed matter physics.
These states play an important role in many unusual phenomena in different materials, like cuprate superconductors, heavy fermions, manganites and others.
In the case of small TM systems, which in modern condensed matter language correspond to nanosystems, clusters and molecules, the role of electron-electron correlations are of special interest from both practical  and fundamental points of view. Indeed, these structures have a great potential to be used in many modern technological applications, like 
ultra-high density magnetic recording systems, sensors and photovoltaic devices. 
From the fundamental point of view,
understanding of correlation effects in small TM systems is important because it will help understand growth dynamics and the role of the quantum effects in the corresponding larger structures. Also, the lack of experimental data for many such small systems places a great burden on theory for predicting their properties.
Unfortunately, most studies of TM clusters either ignore strong electron correlation effects, 
or take them into account in an oversimplified way, thereby neglecting the very
effects that maybe responsible
for their unusual properties such as magnetism and superconductivity.
In the case of systems at the nanoscale correlation effects are expected to be even more important due to reduced dimensionality, which makes it much more difficult for electrons "to avoid" each other as compared
to the case of bulk material.

          Traditionally, in the case of nanostructures and molecules correlation effects are included using 
the DFT+U approach\cite{2,3}. However, in this (mean-field) approximation important dynamical effects 
such as time-resolved local interactions are neglected.
These effects appear to be rather important when correlations are not extremely large (small on-site Coulomb repulsion energy U), 
as in the case of bulk TMs, for which the DFT+U method may lead to  wrong results, such as predicting a non-existing magnetic phase in plutonium.
In the case of large U, the DFT+U approximation may also lead to wrong results, for example
spin-ordering temperatures (see, e.g., review  Ref.~\cite{12} and references therein).

On the other hand, in the case of extended systems several many-body theory approaches have proven to incorporate electron correlation effects successfully. 
Familiar examples include the Bethe ansatz approach in the 1D case (see, e.g., Ref. \cite{5}) and the DMFT method for higher dimensions. In the last approach, one takes into account temporal fluctuations by considering the time- (frequency-) dependence of the electron self-energy. The spatial dependence of the self-energy is neglected,  which strictly speaking is exact  only in the limit of infinite dimensions (coordination number Z)\cite{4}. Nevertheless, 
DMFT appears to be a good approximation for 2D and 3D systems (for an over-review, see Ref.\cite{6}). The DMFT formalism has also been successfully extended
to incorporate electron-phonon interaction \cite{7}, nonhomogeneous systems (multilayered structures) \cite{Potthoff1,Potthoff2,8}, and nonequilibrium phenomena \cite{9}. 

An important step in the analysis of the correlation effects in real materials was made when it was proposed
that the DMFT approach can be incorporated into the DFT scheme \cite{10,11}. In this combined DFT+DMFT method,
the system geometric structure (lattice, interatomic distances etc) and the properties of corresponding ``non-correlated'' (Kohn-Sham) electron subsystem are obtained within the DFT approximations (usually LDA or GGA), 
and correlation effects are taken into account via the short-range Coulomb interaction of the quasiparticles within a Hubbard-type 
tight-binding model. In the case of bulk materials, this approach has been successfully applied for studying the spectral, optical and magnetic properties of systems 
(see Refs.\cite{12} and \cite{13} for reviews). In particular, it was shown that the dynamical fluctuation effects which are naturally incorporated 
in DFT+DMFT approach may lead to a different behavior of the
system from that arising from DFT+U, as mentioned above. 

At first sight, it is not obvious that the DMFT approach can also be applied to nanostructures
or molecules, for which space nonhomogeneity is important in most cases. It has been suggested by Florens\cite{14},
however, that 
when the average atomic coordination number (Z) is large,  DMFT is a good approximation even in the case of finite-sized systems.
Tha application of DMFT to nanostructures, namely, surfaces, nanoparticles, molecules etc, may also be feasible since Z in these systems 
 ranges from vary small to bulk values. The main contribution to the spatial fluctuations
comes from the difference between the ``surface'' (low-coordinated) and the ``bulk'' atoms. Therefore, one can expect
that the spatial fluctuations are important when the number of the surface atoms is comparable to or larger than that of the bulk atoms. This is definitely not the case for nanostructures (except chains), so the DMFT approach must be a reasonable 
approximation for these systems. Moreover, it is expected to be a good approximation even in the case
of much smaller systems, down to non-flat $5-10$ atom clusters, since in this case the average coordination number
is of order of 4, similar to that one of the 2D systems, where DMFT is often considered to be a good approximation. 
Furthermore, there is the generalization of DMFT for the nonhomogeneous case,
in which the electron self-energy remains local, but site-dependent \cite{Potthoff1,Potthoff2}. 
This generalization and enhanced accuracy, which is computationally much more time- and memory-consuming, allows one to study systems with
several hundred atoms within available computational resources (see Section II).

The DMFT and combined DFT+DMFT approximations were recently applied to study several properties of systems containing up to  one hundred atoms \cite{Boukhvalov,Valli,Turkowski,Jacob,Lin,Zgid}.
For example,  spectral properties of the Mn$_{4}$ and H$_{6}$ ring molecules were analyzed
by using DFT+DMFT. \cite{Boukhvalov,Lin} It was shown that the DMFT approximation not only gives resonable results even for such small systems, but also a better
description as compared to the DFT+U \cite{Boukhvalov} or unrestricted HF\cite{Lin} approximations. The transport properties of 110 atom quantum point contact\cite{Valli} and small Ni clusters between Cu nanowires \cite{Jacob} were analyzed
by using DMFT and DFT+DMFT, respectively.
For quantum chemistry, as an alternative to DFT+DMFT a combined Hartree-Fock (HF) and DMFT approximation was used to study the spectral properties of solid bulk hydrogen.\cite{Zgid} The effects of the short range interaction 
were examined within DMFT, while the remaining long-range interactions were treated with the HF approach along these directions. We have recently proposed a combined DFT+DMFT approach to study the magnetic properties
of nanosystems, and showed its validity for examining the magnetic properties of small iron clusters\cite{Turkowski}.

          In this paper, we provide details of our recently proposed combined DFT-DMFT approach to study the physical properties of finite systems, which include nanoclusters and molecules, 
and in which electron-electron correlation play an important role.
We elaborate on the application of the method for nanomagnetism and provide results for the magnetic properties of very small (2-5 atoms) Fe and FePt clusters,
We analyze the role of correlation effects for different cluster geometries and chemical composition. 
Our comparison of the DFT+DMFT and the corresponding DFT+U results suggests that neglect of dynamical correlations tends to lead to an over-estimation of the role of electron correlations. Finally, we discuss possible generalization
of the approach to nanostructures and molecules on substrates and to non-equilibrium systems. 

%V.~Turkowski would like to thank the Organizers of the 51st Sanibel Symposium for the invitation to give a talk and for a kind hospitality. The paper is based on the presentation given to the Symposium.

\section{The DMFT approach: from bulk to nanosize} 

In this Section, for completeness we first review the main equations for the DMFT formalism for extended (homogeneous) systems
and its (nonhomogeneous) generalization for finite systems. 
In most cases, the properties of systems with strong electron-electron correlations can be studied
by solving the problem with the tight-binding Hamiltonian of the Hubbard type:   
\begin{eqnarray}
H=-\sum_{i,j,\sigma ,l,m}t_{il;jm}c_{i\sigma l}^{\dagger}c_{j\sigma m}
+\sum_{i,j,\sigma ,\sigma ',m}U_{\sigma, \sigma '}^{lm}n_{i\sigma l}n_{j\sigma ' m}.
\label{1}
\end{eqnarray}

In the last expression $c_{i\sigma l}^{\dagger}$ and $c_{j\sigma m}$ are the electron creation and annihilation operators of the fermion at site i with quantum numbers $\sigma$ (spin) and 
and l (other quantum numbers: orbital momentum, band number etc). In these notations, the particle number
operator, which corresponds to state $\sigma , l$  is
$n_{i\sigma l}=c_{i\sigma l}^{\dagger}c_{i\sigma l}$.
The competition between the kinetic energy and the (short-range Coulomb repulsion) potential energy is defined
by the hopping  $t_{il;jm}$  and the Coulomb repulsion matrices $U_{i,j,\sigma \sigma '}^{lm}$. 
Usually, the hopping matrix elements are obtained from fitting the experimental band structure,
and only the nearest-neighbor or next-nearest-neighbor hopping matrix elements are taken into account.
The Coulomb repulsion energy is often chosen in the local approximation,  $(U_{i,j,\sigma \sigma '}^{lm}=\delta_{ij}\delta_{lm}U^{l})$, since these terms give the largest contribution into the Hamiltonian in Eq.(\ref{1}),
and the values for $U^{l}$ ($l$ is usually the orbital number: s, p, d ...) are chosen as fitting parameters,
though one can in principle obtain the values for $U^{l}$ from DFT calculations
(see, e.g., Ref.~\cite{13}).

             The problem described by the Hamiltonian in Eq.~(\ref{1}) can be solved by using 
 the time-ordered Green's function (GF)
\begin{eqnarray}
G_{i\sigma l;j\sigma ' m}(t,t')=-i\langle Tc_{i\sigma l}(t)c_{j\sigma m}^{\dagger}(t')\rangle .
\label{3}
\end{eqnarray}
This function is connected to the single-particle self-energy $\Sigma_{i\sigma l;j\sigma ' m}(t,t')$
through the Dyson equation in the following way:
\begin{eqnarray}
G_{i\sigma l;j\sigma ' m}(\omega )=G_{i\sigma l;j\sigma ' m}^{0}(\omega )
+\left[
G^{(0)}(\omega )\Sigma (\omega )G(\omega )
\right]_{i\sigma l;j\sigma ' m}
\label{Dyson}
\end{eqnarray}
(in the frequency representation). In the last equation, $G_{i\sigma l;j\sigma ' m}^{(0)}(\omega )$
is the non-interacting GF (which corresponds to the $U=0$ case).
Since in the  high-dimensional (coordination) limit, the problem can be reduced to the single-site problem with the local GF and the local self-energy, taken for example at sites i=j=0, one can find from the Dyson equation (\ref{Dyson})
the following expression for the local GF:
\begin{eqnarray}
G_{\sigma l;\sigma 'm}(\omega)=\int\frac{d {\bf k}}{(2\pi )^{d}}
\left(
\frac{1}{\omega - \varepsilon ({\bf k})+\mu -\Sigma (\omega )}
\right)_{\sigma l;\sigma 'm}.
\label{4}
\end{eqnarray}
In the last equation, $\varepsilon ({\bf k})$   is the "free" quasi-particle spectrum obtained from the Fourier transform of the hopping matrix, $\mu$  is the chemical potential, and $\Sigma (\omega )$  is the quasi-particle self-energy, which is also local
and site-independent in the case of the DMFT approximation. 
The frequency-dependence of the self-energy allows one to take into account dynamical effects, such as the time-dependent local occupation numbers and interaction. 
In the case of standard mean-field (in particular HF, which corresponds to DFT+U)
approximation, in which $\Sigma$ is  assumed to be static (frequency-independent), these effects are neglected.
Though finding the frequency-dependence of the self energy is often not an easy task, the space-(momentum-) independence of the self-energy is a crucial simplification, which is exact in the limit of infinite dimensions
or coordination number\cite{4}, and allows one to solve the problem of strongly correlated electrons
with rather high accuracy in many cases.
A variety of physical quantities can be obtained from the GF, for example  the orbital spin density
\begin{eqnarray}
n_{\sigma l}=-\int\frac{d \omega}{2\pi }\int\frac{d {\bf k}}{(2\pi )^{d}}{\rm Im} G_{\sigma l;\sigma l}(\omega).
\label{5}
\end{eqnarray}

It follows from Eq.~(\ref{4}) that in order to find the GF one needs to find the self-energy $\Sigma (\omega)$.
Additional equation(s) which connect the local GF and self-energy can be found in the following way.
The local GF can be formally expressed in terms of the path integral over the Grassmann variables $\psi$  and $\psi^{*}$ :
\begin{eqnarray} 
G_{\sigma l;\sigma 'm}(\omega )=\int D[\psi ] D[\psi^{*}]\psi_{\sigma l}\psi_{\sigma 'm}^{*}e^{-A[\psi ,\psi^{*}]},
\label{7}
\end{eqnarray}
where the action  $A$ has the following form:
\begin{eqnarray}
A[\psi ,\psi^{*}]=\int_{0}^{\beta} d\tau \left(
\sum_{i,\sigma}\psi_{i\sigma}^{*}\partial_{\tau}\psi_{i\sigma}
-\sum_{i,j,\sigma}t_{ij}\psi_{i\sigma}^{*}\psi_{j\sigma}
-\mu\sum_{i,\sigma}\psi_{i\sigma}^{*}\psi_{i\sigma}
+U\sum_{i}n_{i\uparrow}n_{i\downarrow}
\right) 
\label{Action}
\end{eqnarray}
(here and in some places below, for simplicity we neglect the orbital degrees of freedom and assume the Coulomb repulsion to be local).
The intergation over all states which correspond to the sites different from $l=0$ in Eq.~(\ref{7}) can be formally
performed, which gives:
\begin{eqnarray} 
G_{\sigma l;\sigma 'm}(\omega )=\int D[\psi ] D[\psi^{*}]\psi_{\sigma l}\psi_{\sigma 'm}^{*}
e^{-A_{eff}[\psi ,\psi^{*},{\cal G}^{-1}]},
\label{impurity}
\end{eqnarray}
where 
\begin{eqnarray}
A_{eff}[\psi ,\psi^{*},{\cal G}^{-1}]=
-\int_{0}^{\beta} d\tau 
\int_{0}^{\beta} d\tau '
\sum_{\sigma}\psi_{0\sigma}^{*}(\tau )
{\cal G}^{-1} (\tau -\tau ')
\psi_{0\sigma}(\tau ')
+U\int_{0}^{\beta} d\tau n_{0\uparrow}(\tau )n_{0\downarrow}(\tau )
\label{Actioneff}
\end{eqnarray}
is the impurity effective action ($\tau$ is the imaginary time and $\beta=1/T$ is the inverse temperature). This action depends on the 
 effective dynamical mean-field
${\cal G}_{\sigma l;\sigma 'm}(\omega)$, which takes into account all effects of the rest of the system
with sites $\not= 0$ on the impurity site. The problem now is equivalent to the problem of the single site coupled
to the bath described by the field ${\cal G}_{\sigma l;\sigma 'm}(\omega)$.
In principle, in the case of given dynamical mean-field one can find the local GF from Eq.~(\ref{impurity})
by using different approaches (see below). 

The next step is to find the dynamical mean-field ${\cal G}_{\sigma l;\sigma 'm}(\omega)$. 
The equation which connects this field with the local GF and self-energy can be found
by mapping the single impurity problem to that of the Anderson impurity with the Hamiltonian:
\begin{eqnarray}
H_{A}=\sum_{\sigma}(\epsilon_{0}-\mu )c_{0\sigma}^{\dagger}c_{0\sigma}
+Un_{0\uparrow}n_{0\downarrow}
+\sum_{{\bf k},\sigma}V_{{\bf k}}\left(
c_{{\bf k},\sigma}^{\dagger}c_{0\sigma}+c_{0\sigma}^{\dagger}c_{{\bf k},\sigma}
\right)
+\sum_{{\bf k},\sigma}\epsilon_{{\bf k}}^{b}
c_{{\bf k},\sigma}^{\dagger}c_{{\bf k},\sigma}.
\label{Anderson}
\end{eqnarray}
where $\epsilon_{0}$ and $\epsilon_{{\bf k}}^{b}$ are the energy levels of the impurity and of the bath electrons,
and $V_{{\bf k}}$ is the hybridization between the impurity and bath states (see, e.g., Ref.~\cite{6}). Since this Hamiltonian is quadratic in the bath-field operators,
one can easily find the impurity Green's function in the non-correlated case (U=0):
\begin{eqnarray}
 {\cal G}(\omega)=\frac{1}{\omega-\epsilon_{0}+\mu -\Delta (\omega)},
\label{MF}
\end{eqnarray}
where 
\begin{eqnarray}
 \Delta (\omega)=\sum_{{\bf k},\sigma}\frac{|V_{{\bf k}}|^{2}}{\omega-\epsilon_{{\bf k}}^{b}},
\label{MF}
\end{eqnarray}
is the hybridizarton function.
Since, we assume that "the Hubbard"  (in Eq.~(\ref{Action})) and "the Anderson" (Eq.~(\ref{Anderson})) baths
are equivalent, the mean-field functions in Eq.~(\ref{Actioneff}) and in Eq.~(\ref{MF}) are equal to each
other (since in both cases they are equal to the local impurity GFs at U=0). On the other hand,
the function (\ref{MF}) is connected with the impurity self-energy in the standard way by means
of the Dyson equation
\begin{eqnarray}
G^{-1}(\omega )={\cal G}^{-1}(\omega )-\Sigma (\omega).
\label{DysonAnderson}
\end{eqnarray} 
Again, since the two impurity problems are equivalent, the self-energies in Eqs.~(\ref{4}) and (\ref{DysonAnderson})
have to be the same.

We thus obtain a closed system of DMFT equations, i.e. Eqs.~ (\ref{4}), (\ref{impurity}) and (\ref{DysonAnderson}),
for the local GF, dynamical mean-field and self-energy. 
It can be solved by iterations for every value of frequency in the following way:
\begin{enumerate}
\item choose the initial self energy $\Sigma (\omega )$;
\item calculate the local GF from Eq.~(\ref{4});
\item calculate the dynamical mean-field ${\cal G}(\omega )$ from Eq.~(\ref{DysonAnderson});
\item use the mean-field function to solve the impurity problem (\ref{impurity}) to find the local GF;
\item find new self-energy from Eq.~(\ref{DysonAnderson}) by using the values of $G (\omega )$
and ${\cal G}(\omega )$;
\item continue iterations until $\Sigma (\omega)$ converges with the desired accuracy.
\end{enumerate}

The most difficult part of the solution is the path integral equation (\ref{impurity}).
Usually, exact numerical methods like Quantum Monte Carlo (QMC) solvers (e.g., Hirsh-Fye or continuous-time QMC approaches), or analytical ones employing expansion either
in powers of t/U or U/t, including the iterative perturbation theory expansion (for details, see Refs.~\cite{6,12,13}), 
are used.

In the case of nanostructures, the problem is essentially nonhomogeneous (except some specific cases,
like symmetric clusters that consist of single type of atoms: dimers, equilateral triangles etc., in which all sites are equivalent),
making the process complex and tedious. To simplify the solution, it was proposed \cite{14}
that  the properties at a given site  (a cavity)
can be obtained by knowing the solution on the rest of the sites. To find the last solution, one can continue by considering the next cavity approximation, by removing the nearest to the cavity shell and so on. 
Thus, when one finds the solution for the external shell one can move backwards, to solve one-by-one the internal shell problems
coupled to the corresponding external shell bath. At the end, one can find the dynamics on the central atom
(see Fig.~1, where the illustration of the reduction of the problem to the cavity problem  for a three-shell nanoparticle is presented).
\begin{figure}[t]
\includegraphics[width=7.5cm]{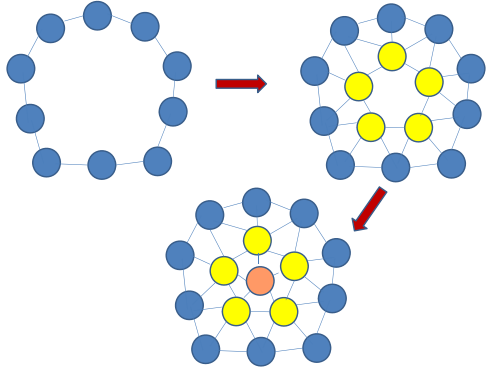}
\caption{\label{fig1} 
The illustration of the reduction of the DMFT problem to the cavity problem in the case of three-shell nanoparticle.
 }
\end{figure}

In a more straightforward manner, in the case of not too large system (up to a few hundred of atoms)
one can solve the problem "exactly" without dividing the system into shells  by generalizing the system of the DMFT equations in real space representation
(the generalization of the DMFT theory on the nonhomogeneous Hubbard model was made by Nolting and Potthoff
\cite{Potthoff1,Potthoff2}).
Namely, the local site-dependent GF for the N-atom system can be obtained from equation (\ref{Dyson}), which
can be written as:
\begin{equation}
{\hat G}^{-1}(\omega )=
\left(
\begin{array}{c}
\omega +\mu-\Sigma_{1}(\omega )\\
t_{21}\\
t_{31}\\
...\\
t_{N1}
\end{array}
\begin{array}{c}
t_{12}\\
\omega +\mu-\Sigma_{2}(\omega )\\
t_{32}\\
...\\
t_{N2}
\end{array}
\begin{array}{c}
t_{13}\\
t_{23}\\
\omega +\mu-\Sigma_{3}(\omega )\\
...\\
t_{N3}
\end{array}
\begin{array}{c}
...\\
...\\
...\\
...\\
...
\end{array}
\begin{array}{c}
t_{1N}\\
t_{2N}\\
t_{3N}\\
...\\
\omega +\mu-\Sigma_{N}(\omega )\\
\end{array}
\right) .
\label{Gnonhom}
\end{equation}
We assume that the self-energy remains local in space, but it is site-dependent:
$\Sigma_{ij}(\omega )=\delta_{ij}\Sigma_{i}(\omega )$.
The other two equations (Eqs.~(\ref{impurity}) and (\ref{DysonAnderson})) 
remain the same, as in the extended case, though one needs to solve N impurity problems
(or less, if some sites are equivalent due to the symmetry of the system):
\begin{eqnarray}
G_{ii}^{-1}(\omega )={\cal G}_{ii}^{-1}(\omega )-\Sigma_{i} (\omega),
\label{Dysonnonhom}
\end{eqnarray} 
\begin{eqnarray} 
G_{ii}(\omega )=\int D[\psi ] D[\psi^{*}]\psi_{i}\psi_{i}^{*}
\exp\left(
-\int_{0}^{\beta} d\tau 
\int_{0}^{\beta} d\tau '
\sum_{\sigma}\psi_{i\sigma}^{*}(\tau )
{\cal G}_{ii}^{-1} (\tau -\tau ')
\psi_{ii\sigma}(\tau ')
\right.
\nonumber \\
\left.
+U\int_{0}^{\beta} d\tau n_{i\uparrow}(\tau )n_{i\downarrow}(\tau )
\right)  .
\label{impuritynonhom}
\end{eqnarray}
 Equation (\ref{impuritynonhom}) can be obtained in the same way as in the bulk case, when instead
of one site one considers every site of the system: $1\leq i\leq N$.
Equation (\ref{Dysonnonhom}) is less obvious. To derive it, one can map the single impurity
problem to the Anderson impurity problem for every site of the system. In this case, one obtains N bath functions ${\cal G}_{ii}(\omega)$. 

Therefore, in the case of nanosystems one needs to solve the system of DMFT equations (\ref{Gnonhom}), (\ref{Dysonnonhom}), and (\ref{impuritynonhom}). 
Similar to the extended case, one can solve the problem by iterative procedure for every value of frequency:

\begin{enumerate}
\item choose the initial self-energies $\Sigma_{i} (\omega )$ for each site;
\item find the local GF, $G_{ii}(\omega)$, from Eq.~(\ref{Gnonhom}) for each site by inverting
         the matrix and extracting the diagonal elements ($i$th diagonal element is equal to $G_{ii}(\omega)$);
\item calculate the mean-fields ${\cal G}_{ii}(\omega )$, $1\leq i\leq N$  from Eq.~(\ref{Dysonnonhom});
\item use the mean-field functions to solve the impurity problems (\ref{impuritynonhom}) to find the local GFs for all sites;
\item find new self-energy from Eq.~(\ref{Dysonnonhom}) by using the values of $G_{ii} (\omega )$
and ${\cal G}_{ii}(\omega )$;
\item continue iterations until $\Sigma_{ii} (\omega)$, $1\leq i\leq N$,  is converged with the desired accuracy.
\end{enumerate}

In both extended and nano-cases, the problem can be solved separately
for every frequency, which makes it easy to parallelize. However, 
the nonhomogeneity in the nano-case makes the calculations far more demanding as
one needs
to work with the inversion of rather large matrices (\ref{Gnonhom}). Assuming that modern
computers are capable of inverting efficiently
matrices of size up to $10.000\times 10.000$, systems with 500 atoms or less 
can be studied by using "the exact" set of equations.
This number is obtained by dividing 10.000 by the numbers of orbitals and spin projections 
which also must be included in the matrix
(\ref{Gnonhom}). Usually, one takes around 6 or 9 orbitals into account, which may include
one s-orbital, three p-orbitals, and five  d-orbitals etc. This allows one to study nanoparticles of size up to $\sim 5nm$.
For larger systems, one could divide them into blocks, solve 
for each block separately, and then sew the solution by solving the problem of the "border" layers between each block, by "freezing"
the internal (core) block sites, i.e. by keeping the local self-energy for these atoms fixed.
The accuracy of the DMFT solution can be estimated by comparing the results with some exact results known
for the Hubbard model. In particular, one can compare the values of the lowest Green's function and self-energy
spectral moments with the exact analytical expressions.\cite{moments1,moments2,moments3}

\section{DFT+DMFT approach: bulk case}

In the case of DFT+DMFT calculations the first step  consists of a DFT study of the system, i.e 
one needs to optimize the system
structure and obtain characteristics of the ``non-correlated'' quasi-particles, such as the bandstructure
(energy $\varepsilon ({\bf k})$ in Eq.~(\ref{4}))) and the density of states (DOS). At the next step, one can extract the parameters for the tight-binding Hamiltonian: the inter-site hopping parameters and short-range (usually, local and nearest-neighbor) 
Coulomb repulsion energies. Next, the correlated problem can be solved allowing estimation of the role of correlations in properties of interest.
To make the analysis more consistent one should perform the DFT calculations by using the DFT+U method,
to ensure that at least the non-dynamical part of the correlation effects is taken into account when optimizing the system geometry and that the band structure is obtained
with the same value of U that is used in the DMFT calculations. However, one must be careful not to take into account the correlation effects twice: in geometry optimization and in the DMFT calculations. The optimization process
is not affected by this double counting, since it is performed only one time, but the band structure
in principle must be corrected by extracting the contribution which comes from the LDA or GGA exchange-correlation (XC) potentials. 
In many cases  this contribution is not very large and may be neglected.

The DFT+DMFT approach continues to be successfully applied to many correlated electron materials.
We refer the reader to excellent reviews of this topic
\cite{12,13}, and to more recent numerous literature, most notably that of iron pnictide superconductors.\cite{Hansmann,Yee}). 

\section{DFT+DMFT approach: nanosystems and molecules}

In the case of small systems, the DFT+DMFT formalism has the same strategy as in the bulk case:
first, to obtain the main properties of "non-correlated" electrons within DFT, or more precisely within DFT+U, and then to use the resulting structures and other quantities to construct and solve the tight-binding Hamiltonian.

Contrary to the bulk, in the nonhomogeneous/nanocase it is more convenient to solve the problem
in real space. Therefore, one needs to extract the effective hopping parameters from the DFT calculation.
In general, the hopping parameters are equal to the matrix elements of the non-interacting
Kohn-Sham Hamiltionian with respect to
the localized atomic-orbital wave functions: 
\begin{equation}
t_{ij\alpha\beta}=\int d{\bf r}\psi_{\alpha}^{*}({\bf r}+{\bf R}_{i})
(-\frac{{\bf \nabla}^{2}}{2m}+V_{atomic}({\bf r}))
\psi_{\beta}({\bf r}+{\bf R}_{i}),
\label{tij}
\end{equation}
where ${\bf R}_{i}$  and ${\bf R}_{j}$ are the site vectors and  $\alpha$ and $\beta$ are the corresponding orbital numbers, and $V_{atomic}({\bf r})$ is the atomic potential. Similar, to the bulk case, the values of the Coulomb repulsion parameter can be obtained either from the DFT calculations or it can be used as a parameter.

Below, before giving a relatively detailed description of our application of this approach to magnetism (Subsection
IV.B), we briefly summarize other applications of the DFT+DMFT approach to study the spectral and transport properties of  nanosystems and molecules.

\subsection{Spectral and transport properties}

Perhaps for the first time the electronic structure of small systems was studied within a DFT(LDA)+DMFT 
by Boukhvalov et al \cite{Boukhvalov}, who used the cluster DMFT approach to study properties of $Mn_{4}$ molecular magnet. It was shown that DMFT predicts correct electronic gap observed experimentally in optical conductivity measurements. The LSDA and LDA+U calculations result in a finite DOS at the Fermi energy and a much smaller gap (0.9eV), correspondingly (in the LDA+U case the same values of Coulomb repulsion and the intra-atomic Hund exchange, U=4eV and J=0.9eV, were used). As these calculations show, the dynamical correlation effects can play a crucial role in the electronic structure of small systems.

The cluster DMFT was also used to analyze the ground state geometry and the on-site projected DOS
of the hydrogen clusters ($H_{4}$ cluster,  $H_{6}$ chain and ring, $H_{50}$ chain) \cite{Lin}. 
As the authors have shown, in the case of intermediate values of Coulomb repulsion
the DMFT is a superior approach comparing to other methods, including the unrestricted HF. In particular, 
DMFT predicts reasonably accurately the correct position of the minimum of the energy of tetragonally-coordinated $H_{4}$ 
cluster as a function of the inter-atomic distance, while the unrestricted HF approach fails in this case. 

The system energy and the DOS of the extended cubic system of hydrogen atoms was studied in Ref.~\cite{Zgid}. In particular, the dependence of the properties of the system on the number of bath orbitals was analyzed, which is an important question
in the case of finite systems. 

In Ref.~\cite{Valli}, the diagrammatic dynamical approach to solve the Hubbard model coupled to a non-interacting environment was used with applications to study the transport properties of different nanosystems, like a few-atom system, benzene molecule and a quantum point contact consisting of 110 atoms. In particular the spectral properties and conductivity of the systems was analyzed by taking into account the possibility of the Mott transition due to the electron correlations.
This approach was proposed as an alternative to avoid using the Cayley-type
of the model, discussed by Florens\cite{14}, where the properties of the central site depend on the properties of the external ones without inverse dependence. Notably, as the authors have shown the n-vertex approximation gives meaningful results already at n=1, i.e. at the DMFT level.

The transport properties of nano-contacts were studied in Ref.~\cite{Jacob}, where the DMFT GFs were used
to calculate the Landauer transmission function. Namely, the case of Ni nanocontacts between Cu nanowires was analyzed
and it was shown that the dynamical correlation effects can significantly alter the behavior of the system.

\subsection{Magnetism}

To solve the nano-DMFT problem for specific Caley-tree-type system, 
one may use Florens algorithm. On the other hand, for small systems (up to few hundred atoms)
we can solve the problem exactly, as outlined above. For systems containing few atoms, one can also use 
DMFT codes for extended (periodic) systems,
by assuming, for example, that every cluster occupies a site on a two-dimensional lattice with a very large lattice constant
such that the cluster-cluster interaction may be neglected (a super-cell approximation). In this case, every atom in the cluster corresponds to the site orbital (which includes the atomic orbitals as an additional quantum number). 
We have applied such an approach
to study the magnetic properties of small Fe and FePt clusters.
The DMFT part of the calculations was performed using the  LISA code \cite{12} with the Hirsch-Fye QMC solver \cite{25}. These calculations were based on the cluster structures  obtained by
a DFT calculations (VASP4.6 code\cite{19}, GGA-PW91 XC potential, for details, see Ref.~\cite{Turkowski}). 
To take correlation effects into account starting at the DFT level, the DFT+U approach was used, where we assumed that
U is site- and orbital-independent (3d- and 4s-states were taken into account for Fe atoms, and 5d- and 6s-states in the case of Pt atom).
 The examples of the cluster geometries used in the DMFT calculations are presented in Fig.2.
As reference value for the Coulomb repulsion we have used U=2.3eV, often employed in the case of bulk Fe.
However, the properties of the system were analyzed by varying U from 0 to approximately 5~eV.
In particular, to check our DFT+U results with experimental data, we compared the bondlength
for the Fe dimer for U=2.3~eV with the experimental value, the only available data for small Fe clusters.
For this case, our calculations find bondlength of $1.99\AA$, with DFT+U 
in good agreement with the experimental estimations $1.87\AA$ 
\cite{20} and $2.02\AA$ \cite{21}. 

From the DFT+U calculations we have obtained values for the hopping parameters for the s- and d-valence electrons
 by using the Slater-Koster matrix approximation\cite{23,24}, in which the effective d-orbital radius $r_{d}$ was approximated by the Muffin-Tin-Orbital values $0.864\AA$ and $1.116\AA$ for the Fe and Pt atoms, respectively (see, e.g., Ref.~\cite{24}). 
In the case of iron clusters, we rescaled the hopping parameters by a factor of 0.367 in order to reproduce the average experimental value for the spin-up and spin-down bandwidth, which are split, in this case.
This rescaling may also be considered equivalent to changing U. 
Indeed, it is not the individual values, rather the ratio between the hopping parameters for the different orbitals for the atoms in question that needs to be kept constant. This also follows from the fact that
all the quantities depend on $U/r_{d}^{3/2}$ and  $U/r_{d}^{3}$, 
as it follows from the renormalization of the Hamiltonian. To make the calculations consistent, we have used the same
renormalization also for the FePt clusters.
The following periodic (cluster) DMFT calculations for the optimized structures
were performed by allowing a weak inter-site (inter-cluster) hopping, for faster convergence. For this purpose,
a weak staggered external magnetic field was also used. The calculations appear to become slow at large values of U and
low temperatures. However, to obtain system properties in the zero electronic temperature limit, one does not need to necessarily consider the case T=0.
When T is much lower than the hopping and Coulomb repulsion parameters - the other energy scales of the system - the properties of the system will be very close to the zero-temperature case.

\subsection{Application of nanoDMFT to small Fe and FePt clusters}

The magnetism of small Fe clusters has been studied by using several approaches, including DFT
(see, e.g.,  Ref.~\cite{15}) and DFT+U (see, for example,  Refs.~\cite{16,17,18}) approximations.
Unfortunately, the experimental data on such small clusters is rather limited\cite{20,21,22}, 
though it is sufficient for estimation of the validity of the DMFT approximation.
In this Subsection, we first summarize the results of our analysis of the magnetic properties of small Fe clusters,
part of which was presented in Ref.~\cite{Turkowski}, and then focus on the properties of bimetallic FePt tetramer.

In the Fe case, we have considered clusters containing 2 to 5 atoms and several possible geometries,
which correspond to the minimum energy configurations at different values of U.
In particular, at U=2.3eV we obtained bondlength of $1.99\AA$ in the case of the dimer (the bulk iron bondlength  is $2.49\AA$). 
Our minimum energy structures also include an  equilateral
triangle, a single-side pyramid for four atoms and a bi-pyramid for five atoms
(Figure 2).
\begin{figure}[t]
\includegraphics[width=15.0cm]{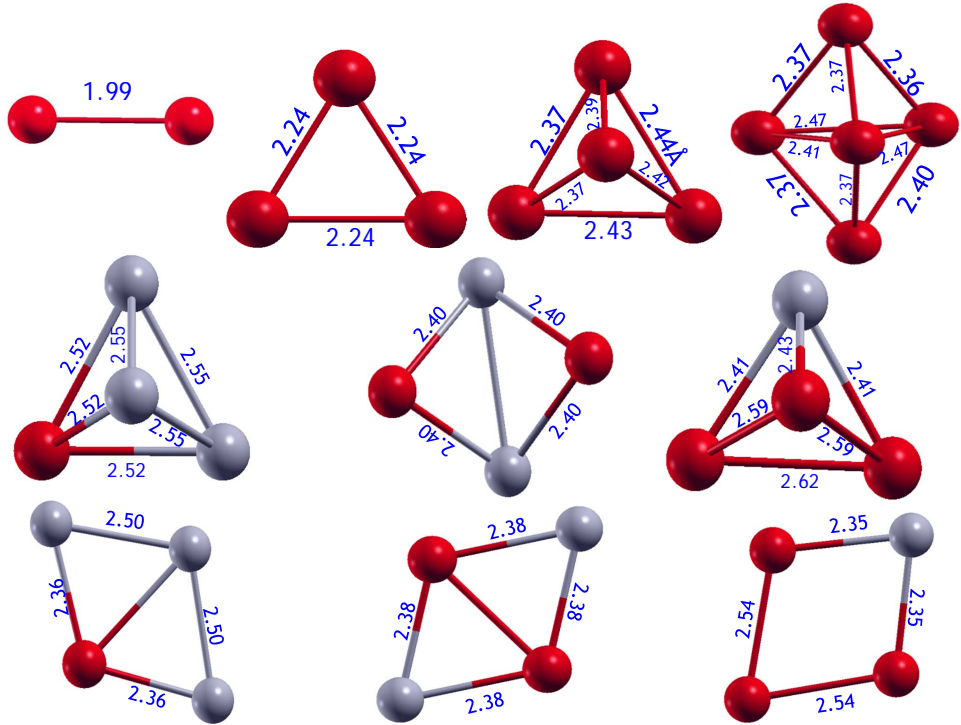}
\caption{\label{fig2} 
Examples of the structures of the Fe (top row, U=2.3eV) and FePt (middle and bottom row, U=4eV) clusters obtained from the DFT+U calculations, that were studied in details in the paper. The red and the gray balls are the Fe and the Pt atoms, correspondingly. The bondlengths are given in Angsroms.  In the case of bimetallic clusters we have presented the lowest energy geometries for both
3D (middle row) and planar (bottom raw) cases.}
\end{figure}
The calculations of the magnetization with the DFT+U approach show that the magnetization per atom
is larger than the bulk value $2.2\mu_{B}$. AS can be seen from the results plotted
in Fig.~3, the magnetization grows as U increases and becomes U-independent after some critical value of the Coulomb repulsion $\sim 1eV$, which depends non-trivially on the number of atoms in the cluster.
\begin{figure}[t]
\includegraphics[width=12.0cm]{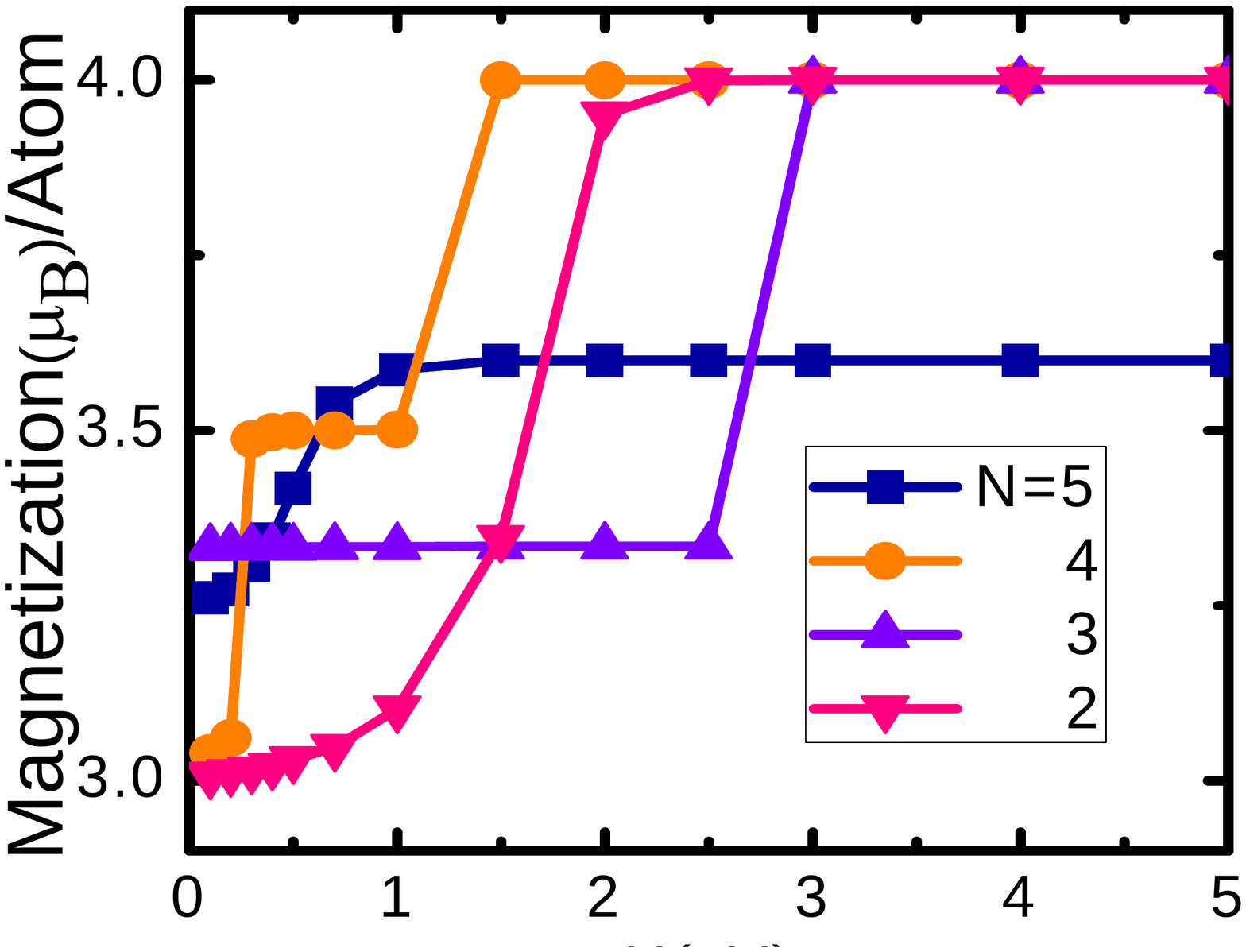}
\caption{\label{fig4} 
DFT+U results for the magnetization as a fuction of U for the $Fe_{2}-Fe_{5}$ clusters.}
\end{figure}

 In  Figure 4, comparison of the results for
$Fe_{3}$ and $Fe_{4}$ obtained with both DFT+U and DFT+DMFT methods are presented, the behaviour is qualitatively the same for clusters of other sizes.
\begin{figure}[t]
\includegraphics[width=12.0cm]{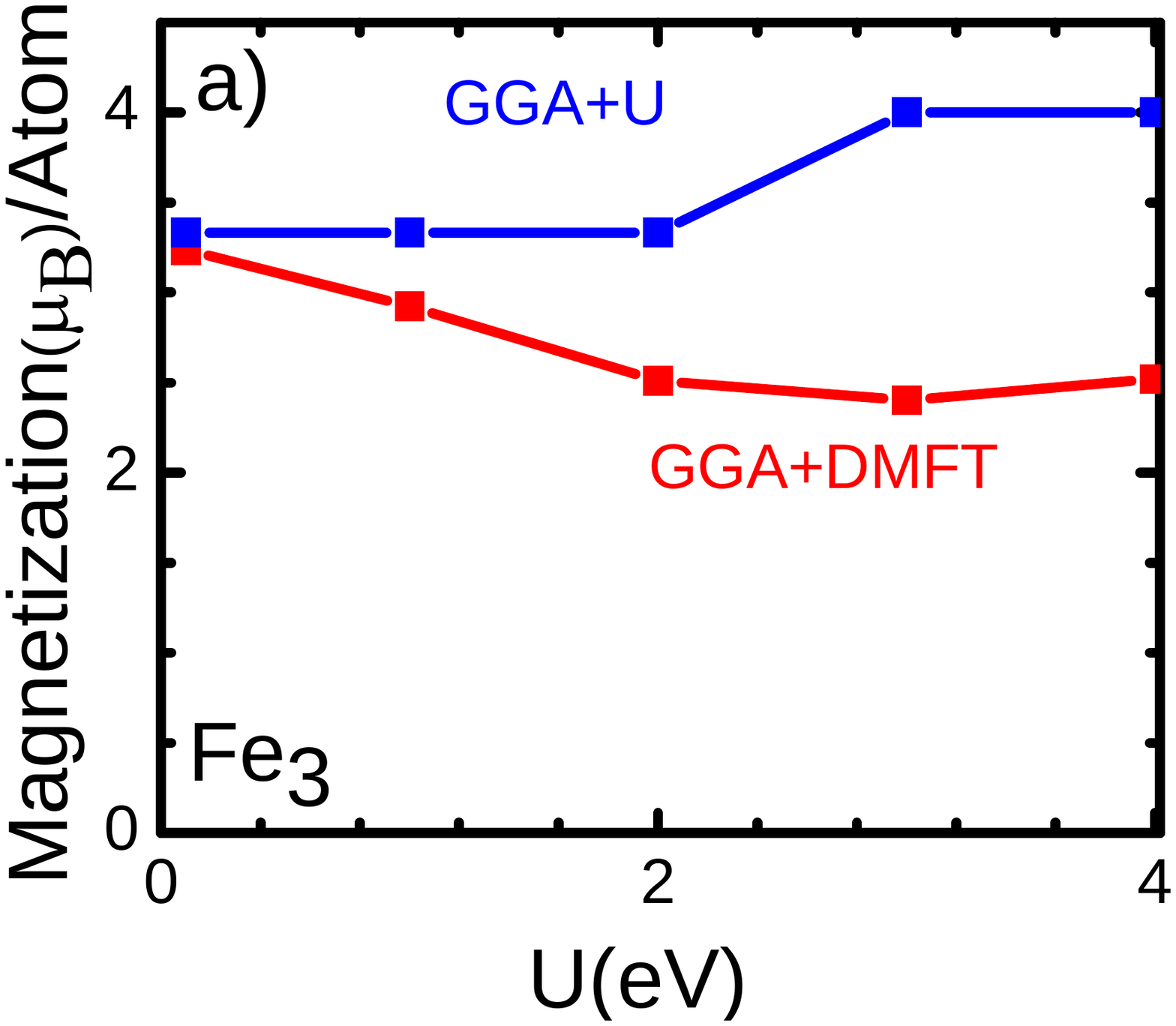}
\includegraphics[width=12.0cm]{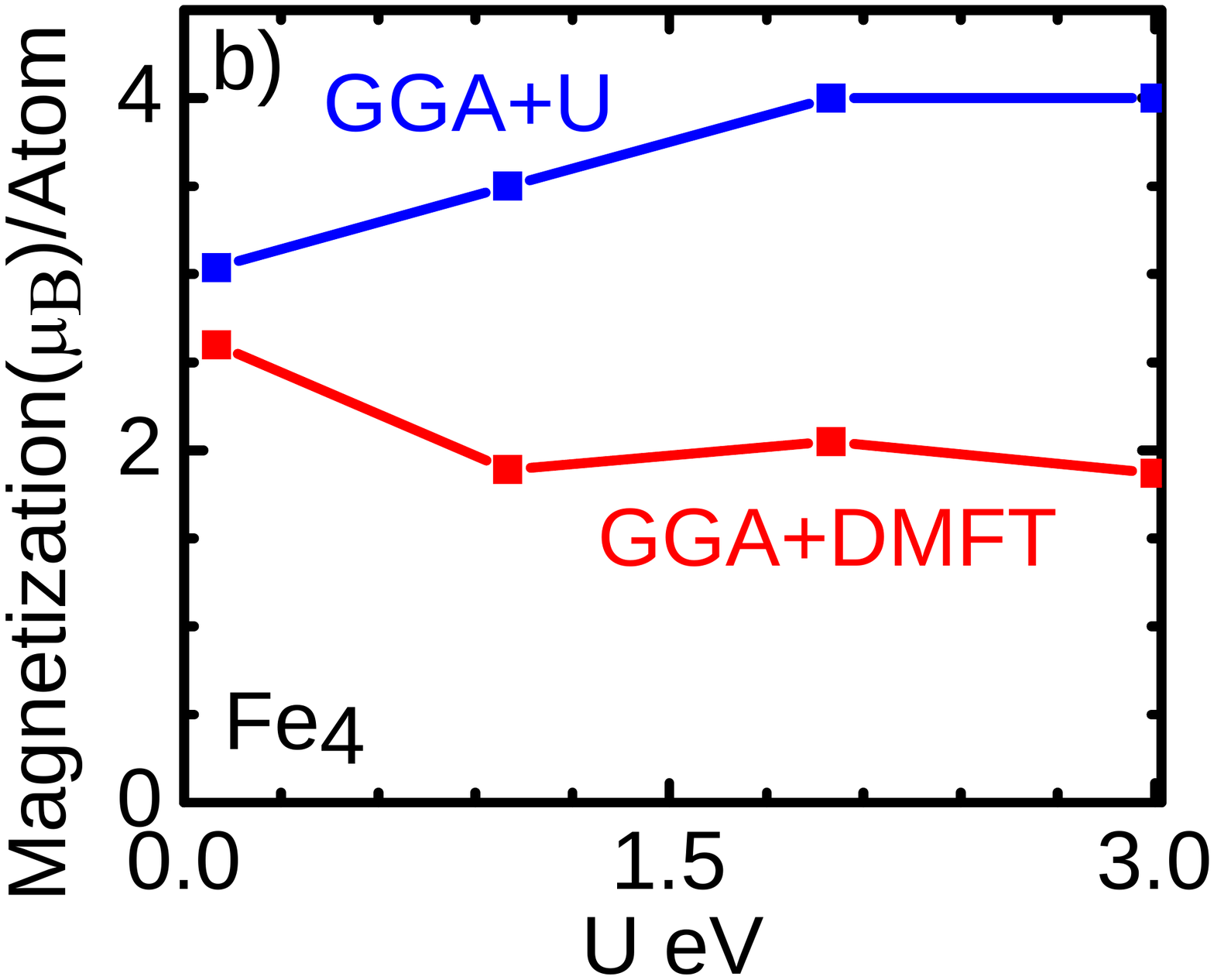}
\caption{\label{fig4} 
DFT+U vs DMFT results for the magnetization of the $Fe_{3}$ and $Fe_{4}$ clusters at different values of U.}
\end{figure}
As it is evident from this Figure, dynamical effects play a role even at rather small values of U,
leading to a significant reduction of the cluster magnetization.

In Table I we present results for cluster magnetization as a function of the number of atoms at characteristic value of U=2.3eV (for more details see Ref.~\cite{Turkowski}).
\begin{table}
\caption{The magnetization per atom as function of the number of atoms in the Fe clusters at U=2.3eV. Along with the DFT+U and DFT+DMFT results we present a set of experimental data \cite{22}.} \label{Table1}
\begin{ruledtabular}
\begin{tabular}{cccc}
N   & DFT+U        &  DMFT         & exp.    \\
 2  &  4.00            &   2.03          &  $3.25\pm 0.5$    \\
 3  &  3.33            &   2.22          &  $2.7\pm 0.33$     \\
 4  &  4.00            &   1.84          &  $2.7\pm 0.8$     \\
 5  &  3.60            &   1.98          &  $3.16\pm 0.33$     \\
\end{tabular}
\end{ruledtabular}
\end{table}
Once again, the DFT+U calculations tend to over-estimate cluster magnetic moments and  dynamical effects in general lead to a reduction of the magnetization. While neither method gives perfect agreement with experiment, DFT+U consistently
overestimates and DFT+DMFT somewhat underestimates the magnetization. The point here was not to obtain
perfect agreement with experiment, rather to show that even for very small clusters, DFT+DMFT gives reasonable  
results.   
One would naturally expect that the DMFT approach  will be much
more accurate in the case of larger clusters, which would make it preferable over DFT+U.
It must be mentioned that the values of magnetization for the selected small clusters used here as a test of our method
are also reasonably well reproduced by other DFT approaches\cite{26,27,28,29},
and most calculated values lie within the experimental error bars, which are quite large.\cite{22}

Finally, for the small Fe clusters we would like to mention briefly our analysis of the the role of the dynamical effects in the case of Jahn-Teller distorted clusters. As our calculations in the case of Fe trimers show, the dependence of the
magnetic moment on the cluster geometry is qualitatively the same in both DFT+U and DMFT cases,
but similar to the previous results DMFT leads to a reduction of the cluster magnetic moment \cite{Turkowski}.
In general such a  significant reduction of the cluster magnetization may indicate 
that the orbital position and/or their occupancy may change dramatically (through the electron self-energy) 
when  dynamical effects are taken into account. 

Theoretical results for the magnetic properties of small FePt clusters is even more limited than those
for the case of Fe.
\cite{Ebert,Boufala}
Since there is hardly any experimental data available, it is very important to perform
a systematic ab initio analysis of such small structures, in particular 
the evolution of their properties with size in order to understand the structure-function relationship.
These clusters
are of special interests for several reasons. For example, there is the need to understand
how the addition of the higher, 5d, localized states of the nonmagnetic metal element
affects the magentic properties of the 3d states of the magnetic atoms in the cluster.
We have chosen Fe- and Pt-atom clusters as our test cases, since such nanoalloys have been the subject 
of both technological and fundamental interest.

In the case of the FePt clusters, we focused on the dimer and the tetramer.
The first case allows one to analyze qualitatively the role of the interplay of the 3d and 5d orbitals in cluster
magnetization. The second case was examined to understand the dependence of magnetism on the chemical
composition of the cluster. Similar to the Fe case, the clusters were optimized by using the DFT+U approach,
but contrary to the iron case it was found that the geometry of the clusters strongly depends on the value of U (Figure 2). In some cases two-dimensional (2D)
structures have lower energy. In particular, the relative energies for the corresponding structures with the same values of Fe and Pt atoms
at U=4eV are: $Fe_{1}Pt_{3}$ - planar: -0.13eV, 3D: 0eV;  $Fe_{2}Pt_{2}$ - planar: -0.63eV, 3D: 0eV; $Fe_{3}Pt_{1}$ - planar: 0eV, 3D: -0.13eV.
The average bond length is longer in the 2D geometry as compared
to the 3D ones, which would make correlation effects more important for them. 
This follows from the notion that one may assume that in the 2D case less orbitals
are involved in the hybridization, and thus correlation effects are enhanced because of the contribution of unbonded charge.

Since the cluster geometry and the energy levels used in the DMFT analysis are obtained with the DFT+U calculations,
we first analyzed how these quantities are affected by changing U.
 The DFT+U results for the magnetization of $Fe$ and FePt dimers as function of U
presented in Figure 5,
 \begin{figure}[t]
\includegraphics[width=12.0cm]{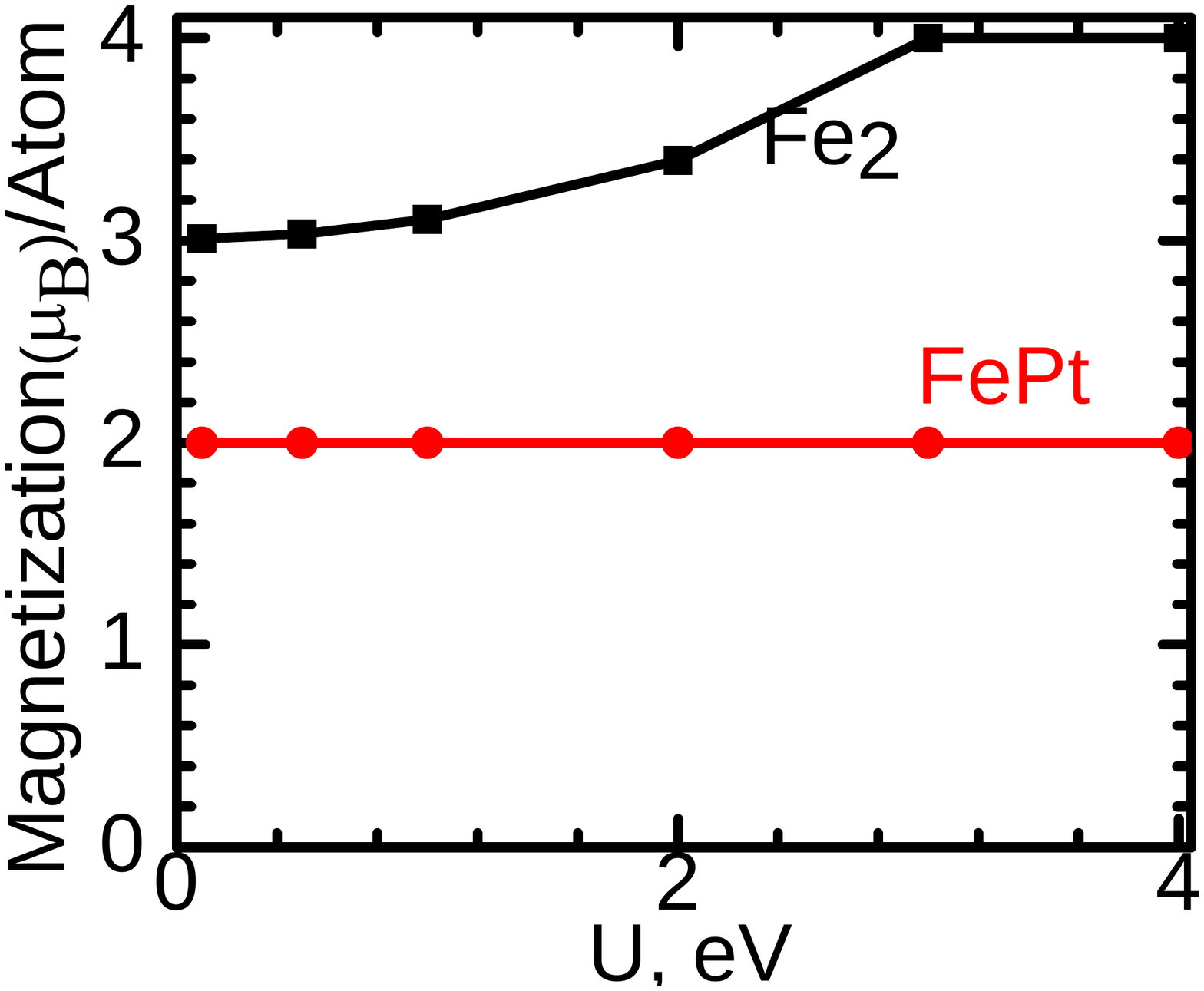}
\caption{\label{fig5} 
DFT+U results for the magnetization of the $Fe_{2}$ and $FePt$ clusters at different values of U.}
\end{figure}
show that the magnetization barely depends on U for the latter
(the value of the magnetization $2\mu_B$ per atom is in agreement with other DFT calculations\cite{Boufala}),
but it has a sharp increase at $U\sim 3eV$, in the case of the former. This result can be explained as follows.
Since increase of U leads to an increase of the
bondlength, and this value of U is of order 3eV - a typical value which corresponds to the Mott transition in bulk materials, and the change of the cluster magnetization is rather sharp at this point, one may speculate the presence of some kind of (magnetic) transition. The fact that the magnetization is U-independent in the case of 
FePt cluster can be explained by assuming that the d-orbitals of the FePt are much less hybridized in the last case 
as compared to the Fe dimer due to the energy mismatch even at small U, so increasing U which leads to an increase of the bondlength does not lead to a significant charge redistribution (localization) as compared to the case of the iron dimer.
In fact, our DFT+U calculations give the following results for the bondlength (at U=2.3eV): $1.99\AA$ ($Fe_{2}$) and 
$2.29\AA$ (FePt). In the case of increasing U the bondlength increases in both cases, for example  in the case of FePt
the bondlength is equal to $2.34\AA$ at U=4.0eV (In the case U=0eV, the DFT calculations\cite{Boufala} give the bondlength $2.2\AA$).  

To get a better understanding of the role of the correlation effects in the spectral properties of the dimers we have calculated the positions of the s- d-state peaks as functions of U for both clusters (Fig.6). 
\begin{figure}[t]
\includegraphics[width=12.0cm]{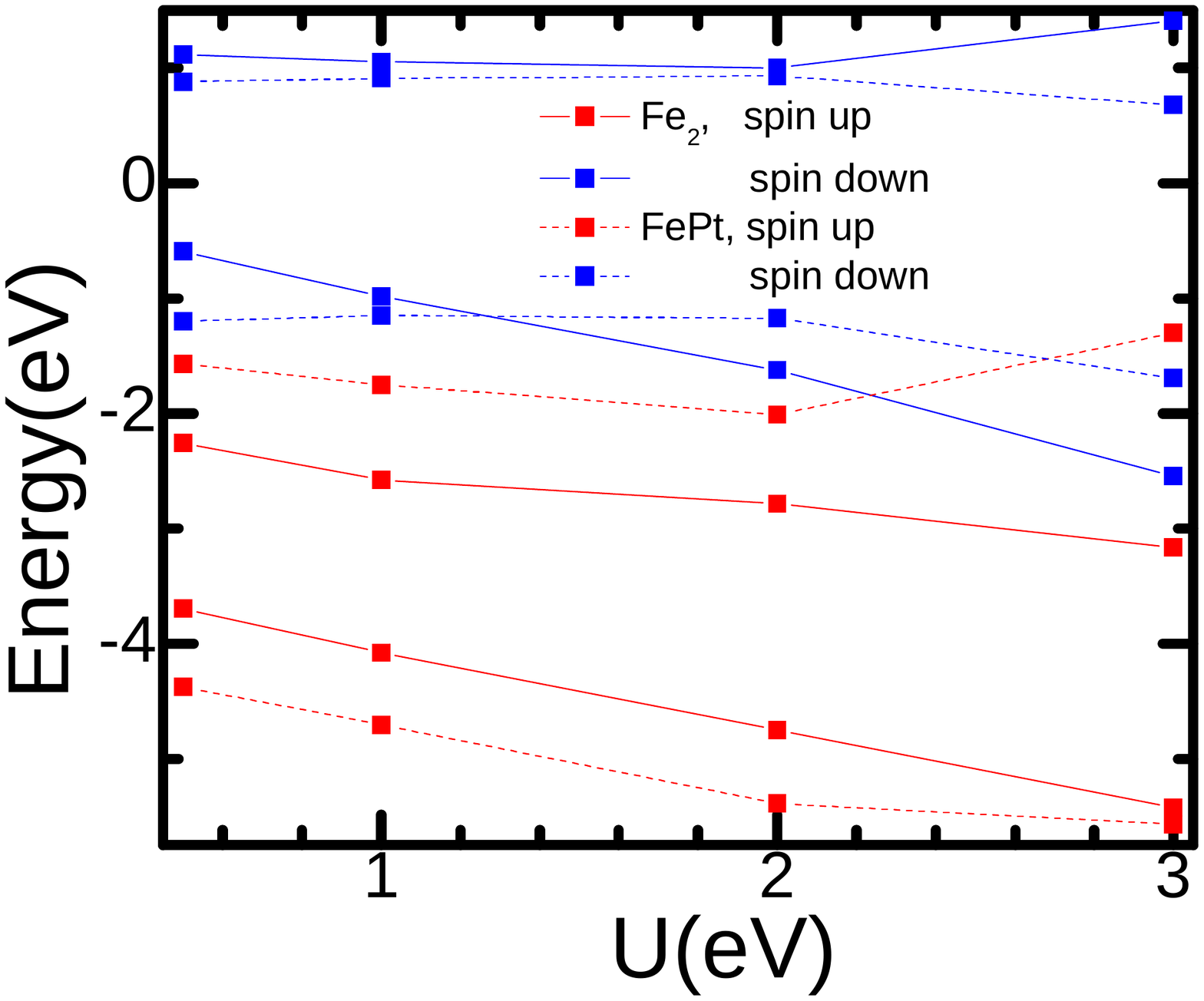}
\caption{\label{fig6} 
DFT+U results for the positions of the s- and d-energy peaks in the DOS of the $Fe_{2}$ and $FePt$ clusters different values of U.
The blue and red solid lines correspond spin up and spin down peaks of the $Fe_{2}$ cluster.
The dotted line correspond to the energy peaks for the FePt cluster.
}
\end{figure}
We find that
with U increasing the distance between the spin-down peaks (two red lines) and spin-up d-orbital peaks (top blue solid line) significantly increases when U changes from 2eV to 3eV in the case of iron dimer. This implies that the the probability for one electron to migrate back and fourth from the spin up
to the spin down orbital, and hence to decrease the cluster magnetization, decreases at large Us. This transition
may correspond to the magnetization jump in Fig.5. In the case of FePt cluster, there is no such sharp change
between the energy difference for these states (the corresponding dashed lines). Therefore, the magnetization 
does not change significantly with U. There is another interesting effect of changing U in the case
of FePt cluster. Namely, the positions of the spin up d-state and spin-down s-state change when U is between 2eV and 3eV.
This means that at large Us one can neglect to some extend the role of s-orbital in correlation effects for FePt cluster, since this orbital may be assumed to be doubly occupied (this statement is true only when the distance between the s-spin-down and d-spin-up peaks is large enough, though).

To obtain a deeper understanding of the role of different d-orbitals in the dynamical processes, we have analyzed the 
positions of projected d-states at different values of U for the FePt dimer (Figs.7,8). It follows from
these Figs. that with increasing U the position of the spin-down $d_{x^{2}-y^{2}}$ state changes significantly with respect
to the Fermi energy (0eV), while that of other orbitals (or the position of the HOMO-LUMO gap) almost do not change, which means that their occupancies are not very much affected by U.
The other important consequence of increasing U is an increase of the distance between the s- and d- states of the same spin,
especially remarkable for the spin-up states. 
\begin{figure}[t]
\includegraphics[width=12.0cm]{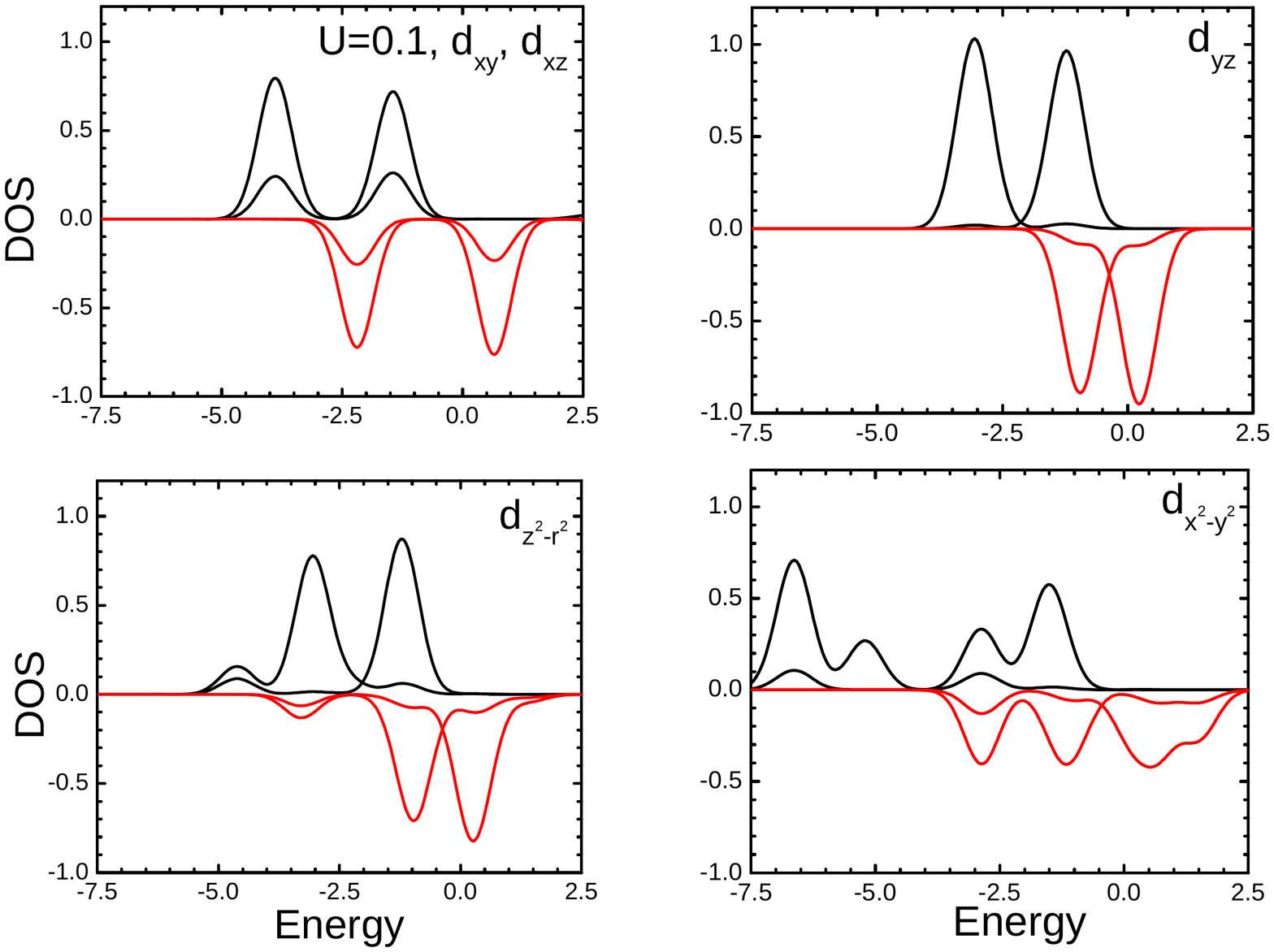}
\caption{\label{fig7} 
The density of states of different d-orbitals obtained within the DFT+U approach at U=0.1eV.}
\end{figure}
\begin{figure}[t]
\includegraphics[width=12.0cm]{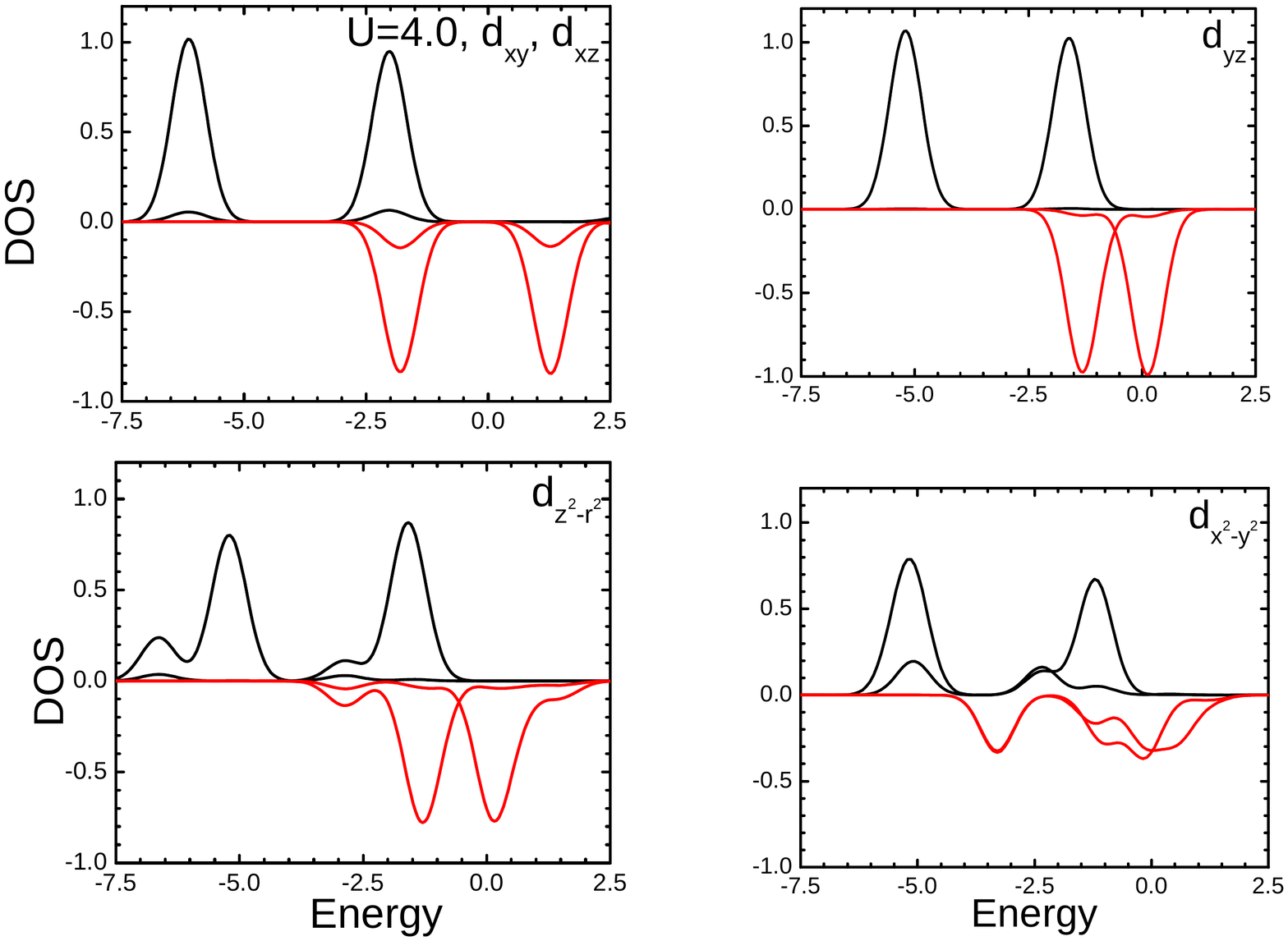}
\caption{\label{fig8} 
The same as in Fig.7 at U=4.0eV.}
\end{figure}
This separations indicates that one can neglect low-lying
spin-up states for the s-orbitals for large U, except the  $d_{x^{2}-y^{2}}$ orbital (Figure 8). In other words, one can neglect the hopping processes between these and other states. In many cases, such a simplification
may lead to a significant reduction of the computational time in the DMFT calculations.

We have also anayzed the magnetic properties of four atom FePt systems.
In particular, we have studied the U-dependence of the magnetization of the 3D and planar $Fe_{3}Pt$
by using both the DFT+U and DFT+DMFT calculations (Figure 9).
\begin{figure}[t]
\includegraphics[width=12.0cm]{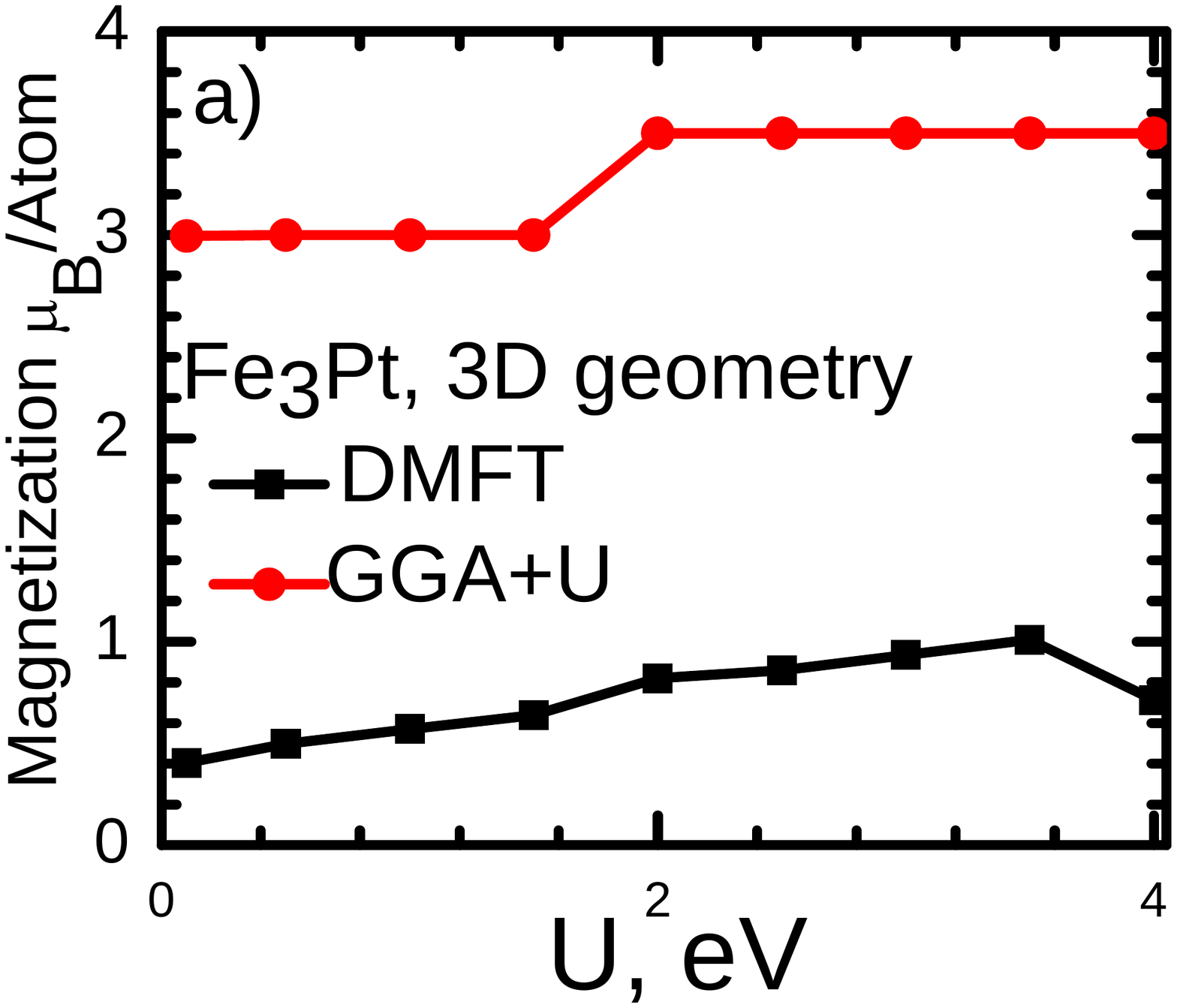}
\includegraphics[width=12.0cm]{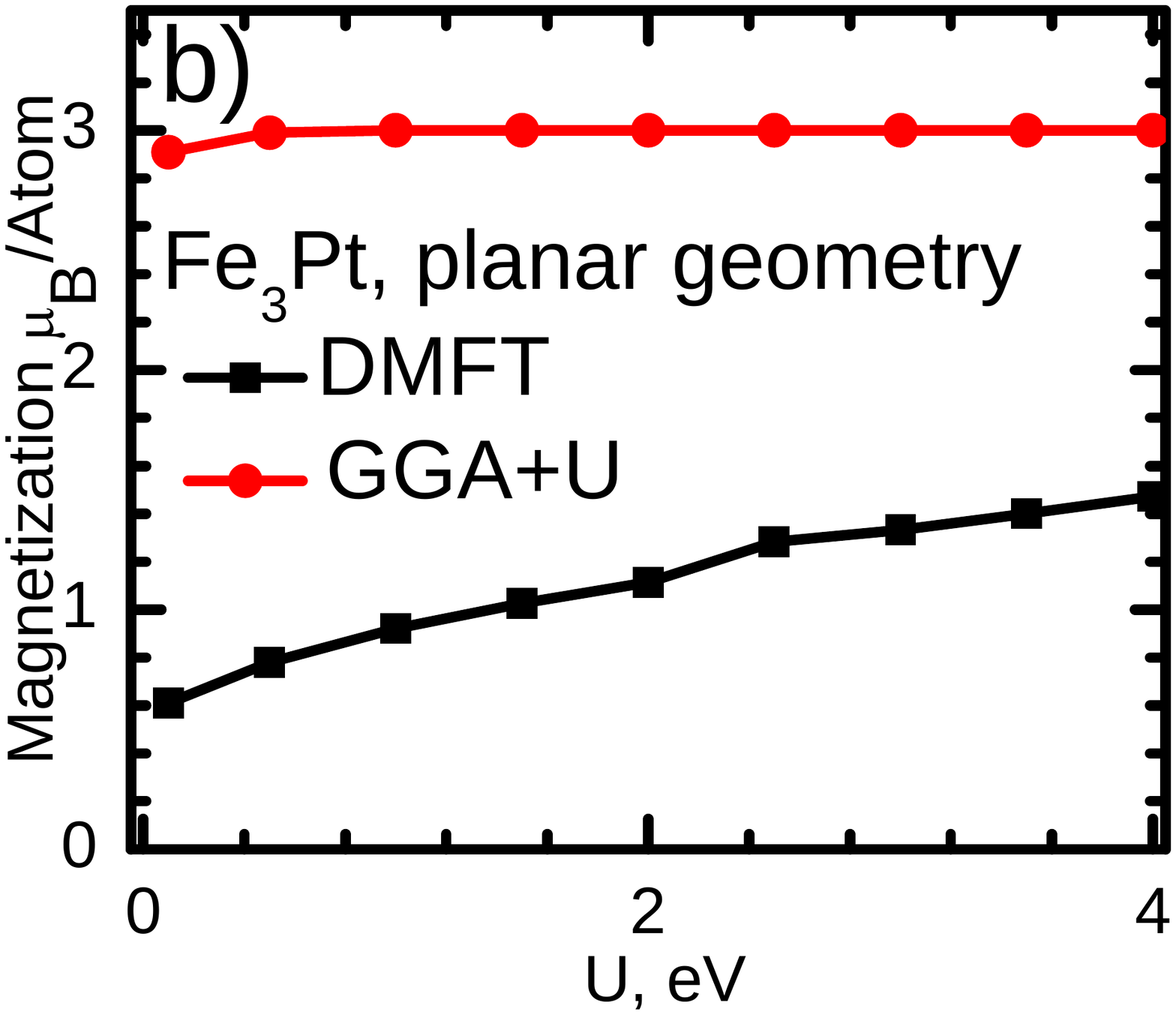}
\caption{\label{fig9} 
DFT+U vs DMFT results for the magnetization of the $Fe_{3}Pt$ clusters as a function of U
in the case of 3D (a) and planar (b) geometries.}
\end{figure}
From Fig.~9 it follows that DFT+U leads to an increase of magnetization at large values of U for
the 3D structures. This is not surprising, since in the case of 3D structure the correlation effects are more sensitive to the distance between atoms than in the 2D case in which the atoms are on average much further from each other 
even for small values of U.  The reduction of magnetization due to dynamical effects, taken into account
in the DMFT is even more dramatic than in the case of pure iron four-atom cluster (Fig.4b). Indeed, when one iron atom is substituted by a platinum atom, the corresponding hopping parameter to this size increases significantly due to larger
spatial extension of the corresponding orbitals, which decreases the average in-time occupancy of the orbitals,
and hence decreases the magnetization.

The chemical composition dependence of the magnetization for the four-atom clusters is presented in Fig.10. 
\begin{figure}[t]
\includegraphics[width=12.0cm]{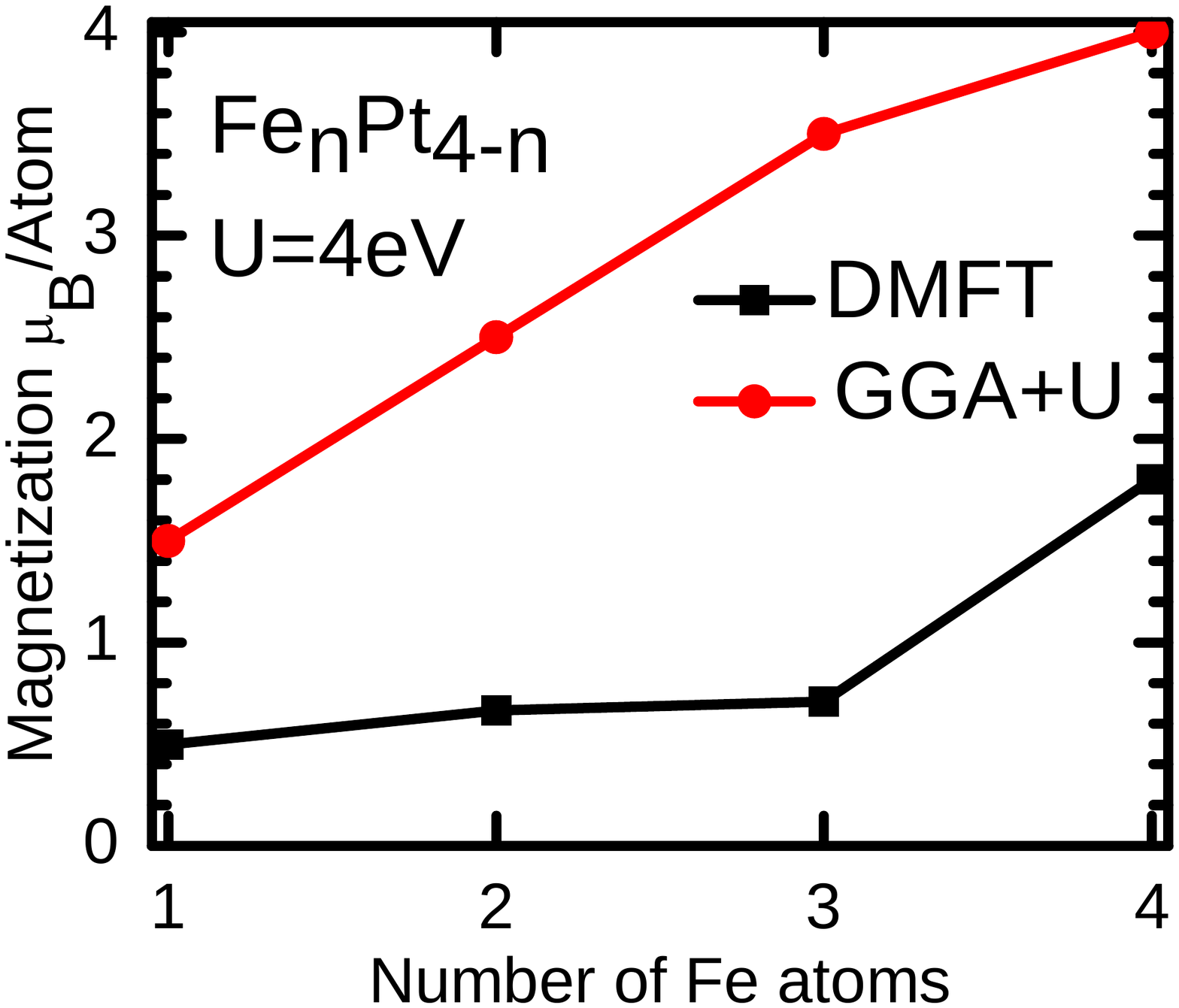}
\caption{\label{fig10} 
DFT+U vs DMFT results for the magnetization of the four atom iron-platinum clusters as function
of the number of iron atoms at U=4eV.}
\end{figure}
In this case, the magnetization increases dramatically with increasing of the number of "more magnetic" Fe atoms
as the DFT+U calculations show. The DMFT calculations show that that the magnetization is rather small  and
almost composition-independent when Pt atoms are present. This means that the hybridization effects
are dominant even in the case of large Us. 
It is interesting to note that the DFT calculations for the $Fe_{2}Pt_{2}$ cluster \cite{Boufala}
give the magnetization $2\mu_{B}$, which is much closer to our DFT+U result than to the DMFT one.
This suggests that the energy levels calculated by the DFT approach are well-separated
already at small U, and the hybridization is rather weak in this case. On the other hand, the dynamical effects
may change dramatically the magnetization even in such a case of strongly-localized orbitals.

\section{Possible extensions}

Since in most practical applications the clusters are supposed to be on a substrate, it would be very
helpful to develop a DFT+DMFT formalism for such cases. One may think about this situation
as a cluster coupled to a bath (substrate). Thus, for the lowest-order approximation
one can solve the problem for the surface and than to take the substrate-cluster interaction into account
by fixing the substrate energy levels. The inverse influence of the cluster on the substrate can be also
taken into account as the next step, similarly to the approach proposed by Florens\cite{14}.

Another important case is the nonequilibrium behavior of nanosystems with strong electron-electron correlations.
In many technological applications, like electronic and magnetic devices, usually fast  switching of strong
external fields takes place. Such a nonequilibrium formulation of DMFT was already done in the case of extended systems\cite{9}.
In this case, one can study the time evolution of the local GF defined on the Kadanoff-Baym or Keldysh
time contour. This problem is technically much more complicated than the time translationally-invariant case, since instead of solving one DMFT
problem for each frequency in the equilibrium case, one needs to work with with complex GF and self-energy
time matrices of a rather large size. The size of the GF matrix will be (number of atoms$\times$number of orbitals)$\times$(number of time points) instead of (number of atoms$\times$number of orbitals) as in Eq.~(\ref{Gnonhom}). 
Moreover, the number of time points must be twice the number of the real time points
in the case of Keldysh contour, and the last number plus the number of (imaginary branch) points on the interval [0,1/T]
for the temperature averaging in the case of Kadanoff-Baym contour. Typically, the number of the real time points
in the nonequilibrium problem
is around 1000 for the real branch and 10-100 for the imaginary one. Obviously, such large matrices
can be barely inverted when one would like to solve the problem exactly, except the case of a few-atom cluster.
In the case of nanosystems, one needs to use again the layer-by-layer solution proposed by Florens.
Presently, we are working on developing numerical tools 
which will allow us to examine the magnetic characteristics of nanoalloys on substrates and also in presence of external fields.

\section{Conclusion}

In this paper, we have provided and overview of the current status of the applications
of combined DFT+DMFT approaches to study correlation effects in the case of molecules and
nanostructures. In general, all the studies show that similar to the case of bulk and 2D systems, this approach is a promising method to take into account the correlation effects even in the case of small clusters and molecules.

We have paid special attention to the case of nanomagnetism, for which we have proposed 
a DFT+DMFT approach and applied it to study the magnetic properties of small Fe and FePt clusters.
It follows from our calculations that dynamical correlation effects lead to a significant decrease of the magnetization than DFT+U. There are two reasons for this: first, the time-dependence of the orbital occupancy
taken into account in the DMFT approach leads to a significant decrease of the magnetization relative to "the staggered"
DFT+U case of "frozen" spins. Second, the frequency-dependence of the self-energy can lead to a shift of the energy levels with respect to the Fermi energy, and hence to a change of the level occupancy and magnetization.
We have analyzed also the geometry, chemical composition and local Coulomb repulsion dependencies of the magnetization
in the case of different clusters. In particular, we find that for a given system, to get the same magnetization
from DFT+U and DMFT one needs to use a larger value of U in the latter, similar to results for the bulk.
One of the reasons for this may be a stronger screening in the DMFT case\cite{12}.
The DFT+U and DFT+DMFT results are approximately on the same level of agreement with the experimental data. However, in the case of larger systems, one may expect that the DMFT results will much more
accurate than the DFT+U.

We have discussed a possible extension of the approach on two most important cases: particles and molecules on a substrate and nonequilibrium systems. The work in these two directions, including application to larger clusters, is in progress.

\section*{Acknowledgements}

We would like to thank DOE for a financial support under Grant No. DOE Grant DE-FG02-07ER46354.

\end{document}